\newcount\subeqnno
\def\bse{\begingroup
\refstepcounter{equation}
\subeqnno=\arabic{equation}
\setcounter{equation}{0}
\def\theequation{\the\subeqnno\alph{equation}}}
\def\ese{\setcounter{equation}{\the\subeqnno}\endgroup}

\newcommand{\mb}[1]{\ifmmode#1\else\mbox{$#1$}\fi}

\newcommand\al{\mb{\alpha}}

\newcommand\ga{\mb{\gamma}}
\newcommand\de{\mb{\delta}}

\newcommand\la{\mb{\lambda}}
\newcommand\si{\mb{\sigma}}

\newcommand\calD{\mb{{\cal D}}}

\newcommand\calG{\mb{{\cal G}}}
\newcommand\calH{\mb{{\cal H}}}

\newcommand\calL{\mb{{\cal L}}}
\newcommand\calM{\mb{{\cal M}}}
\newcommand\calN{\mb{{\cal N}}}

\newcommand{\beq}{\begin{equation}}
\newcommand{\eeq}{\end{equation}}
\newcommand{\nn}{\nonumber}
\newcommand{\bea}{\begin{eqnarray}}
\newcommand{\eea}{\end{eqnarray}}
\newcommand{\norm}[1]{\parallel \! {#1} \! \parallel}

\newcommand{\inprod}[2]{\langle {#1}, {#2} \rangle}
\newcommand{\inproda}[2]{\{ {#1}, {#2} \}}

\newcommand{\deriv}[2]{\frac{d {#1}}{d {#2}}}

\newcommand{\x}{\mb{\times}}
\newcommand{\rhat}{\mb{\underline{\hat{r}}}}

\newcommand{\Ad}{\mb{{\rm Ad}}}

\newcommand{\tr}{\mb{{\rm tr}}}

\newcommand{\ul}[1]{{\underline{#1}}}

\documentstyle[12pt]{article}
\topmargin = - 0.5 cm
\begin{document}
\bibliographystyle{unsrt}

%theorem structures
\newtheorem{result}{Result}
\newtheorem{definition}{Definition}

%the text input

%%%%%%%%%%%%       abstract and title page    %%%%%%%%%%%%%%%%%%%%%%

\begin{flushright}
Imperial/TP/98-99/26\\
DAMTP-1999-41 
\end{flushright}
\begin{center}
\LARGE{Electroweak Vacuum Geometry} \\
\vspace*{0.5cm}
\large{
Nathan\  F.\  Lepora
\footnote{e-mail: N.F.Lepora@ic.ac.uk}$^{,1,2}$
and T. W. B. Kibble
\footnote{e-mail: T.Kibble@ic.ac.uk}$^{,1}$\\}
\vspace*{0.2cm}
{\small\em 1) Blackett Laboratory, Imperial College,\\
Prince Consort Road, London, SW7 2BZ, England\\ }
\vspace*{0.2cm}
{\small\em 2) King's College, Cambridge University\\
Cambridge, CB2 1ST, England\\} 
\vspace*{0.2cm}

{February 1999}
\end{center}

\begin{abstract}
We analyse symmetry breaking in the Weinberg-Salam model paying
particular attention to the underlying geometry of the theory. In this
context we find two natural metrics upon the vacuum manifold: an 
isotropic metric associated with the scalar sector, and a {\em
squashed} metric associated with the gauge sector. Physically, the
interplay between these metrics gives rise to many of the
non-perturbative features of Weinberg-Salam theory. 
\end{abstract}

\thispagestyle{empty}
\newpage
\setcounter{page}{1}

%%%%%%%%%%%%%%%%%%%%%%%%%%%%% section 1        %%%%%%%%%%%%%%%%%%%

\section{Introduction}

Numerous experiments have shown that perturbative Electroweak
interactions are
described by the Weinberg-Salam theory of a broken isospin-hypercharge
gauge symmetry. Within that theory the vacuum is
instrumental in breaking the isospin-hypercharge $SU(2)_I \x U(1)_Y$ 
symmetry to a residual electromagnetic $U(1)_Q$ theory. Physically,
the symmetry is broken by the coupling between this vacuum and
isospin-hypercharge gauge fields. This coupling induces mass for the
W and Z components of the gauge fields, whilst the photon does not
couple, remains massless and represents the residual gauge theory. 

However, in describing the above symmetry breaking Weinberg-Salam
theory also makes definite predictions about the vacuum
structure. Owing to the $SU(2)_I \x U(1)_Y$ gauge symmetry, vacua are
predicted to be degenerate and collectively form the vacuum {\em
manifold}. This structure then implies the existence of
non-perturbative solutions, for example the electroweak strings and
the sphaleron. Thus in describing electromagnetism in
conjunction with a W and Z sector, Weinberg-Salam theory {\em implies}
the existence of further non-perturbative solutions related to the
vacuum structure.

By explicit calculation the relevant vacuum manifold $M$ is predicted
to be a three-sphere. This three-sphere is related to the gauge
structure by the familiar relation
\beq
\label{M}
M = S^3 \cong \frac{SU(2)_I \x U(1)_Y}{U(1)_Q}.
\eeq
Now a reasonable question is: what is special about this relation? For
instance how does $SU(2)_I \x U(1)_Y/U(1)_Q$ differ from
$SO(4)/SO(3)$, which is also isomorphic to a three sphere. 

Mathematically, the answer to this question is: $SO(4)/SO(3)$ and
$SU(2)_I \x U(1)_Y/U(1)_Q$ describe different {\em metrical
structures} on the three-sphere. Both describe homogenous metrics, but
they differ in the symmetry properties of the metric. Essentially
$SO(4)/SO(3)$ is associated with a homogenous and isotropic metric on
the three-sphere, whilst the other gives a homogenous and {\em
anisotropic} metric. 

Within this paper we examine how this mathematical structure relates
to the vacuum structure. It turns out that the anisotropic $SU(2)_I \x
U(1)_Y/U(1)_Q$ metric is naturally induced by the gauge 
sector, with the degree of anisotropy specified by the gauge coupling
constants. In addition the isotropic $SO(4)/SO(3)$ metric is
naturally induced by the scalar sector of Weinberg-Salam theory.
Thus Eq.~(\ref{M}) is interpreted as appertaining to the gauge sector
geometry of Weinberg-Salam theory, whilst $S^3 \cong SO(4)/SO(3)$
appertains to the scalar sector.

We then use this framework to interpret the non-perturbative solution
spectrum in terms of the vacuum geometry. Such solutions are usually
specified in terms of their boundary conditions on the vacuum
manifold. Electroweak strings define embedded circles, whilst
the sphaleron defines an embedded two-sphere. With respect to the
geometry we show that these boundary conditions relate to totally
geodesic submanifolds of the three-sphere with respect to {\em both}
the scalar and gauge metrics. We also show that the scattering of
electroweak strings relate to the holonomy of their boundary
conditions with respect to these metrics. 

\section{Electroweak Symmetry Breaking}

We start by quickly running through the symmetry breaking mechanism in
electroweak theory. We use this discussion to make explicit some of 
the mathematical structure required for this paper. 

The usual approach is taken, whereby a Lagrangian describes the
interaction of the gauge fields $W^\mu \in su(2)$ and $Y^\mu \in
u(1)$ with a scalar field $\Phi \in {\bf C}^2$. Minimisation of the 
scalar potential yields a vacuum with less gauge invariance than the
original symmetry. Then orthogonal rotation of the $W^\mu$ and
$Y^\mu$ fields gives the $W, Z$ and photon basis of mass eigenstates
with respect to this vacuum. 

The scalar-gauge sector of Weinberg-Salam theory is described by the
Lagrangian    
\beq
\label{lag}
\calL = -\frac{1}{4} \inprod{W^{\mu \nu} + Y^{\mu \nu}}{W_{\mu \nu} +
Y_{\mu \nu}} +
(\calD^\mu \Phi)^\dagger (\calD_\mu \Phi)
- \la(\Phi^\dagger \Phi - v^2)^2,
\eeq
with the field tensors 
\bse
\bea
W^{\mu \nu} &=& \partial^\mu W^\nu - \partial^\nu W^\mu + [W^\mu,
W^\nu],\\
Y^{\mu \nu} &=& \partial^\mu Y^\nu - \partial^\nu Y^\mu,
\eea
\ese
the covariant derivative 
\beq
\calD^\mu = \partial^\mu + W^\mu + Y^\mu,\\
\eeq 
and $\inprod{\cdot}{\cdot}$ the $su(2) \oplus u(1)$ inner product.

There is some freedom in the choice of non-degenerate inner products
$\inprod{\cdot}{\cdot}$ on $su(2)_I \oplus u(1)_Y$, constrained to be
invariant under the adjoint action of $SU(2)_I \x U(1)_Y$. We
parameterise the possible inner products in the following way 
\beq
\label{inprod}
\inprod{X}{Y} = -\frac{1}{g^2} ( 2\tr XY - \tr X \tr Y )
- \frac{1}{g'^2} \tr X \tr Y,
\eeq
with real, positive parameters $g$ and $g'$. Choosing a basis
$\{\frac{1}{2}i \si_a\}$ for $su(2)$, with $\si_a$ the Pauli spin
matrices, and $\frac{1}{2} i {\bf 1}_2$ for $u(1)$, we see that the
unit norm generators are  
\beq
\norm{\frac{1}{2}ig\si_a}=\norm{\frac{1}{2}ig'{\bf 1}_2}=1.
\eeq
With these generators the covariant derivative explicitly takes the
form
\beq
\calD^\mu = \partial^\mu + \frac{1}{2} ig \si_a W^\mu_a
+ \frac{1}{2} ig' Y^\mu,
\eeq 
yielding the familiar interpretation of $g$ and $g'$ as the isospin
and hypercharge gauge coupling constants.

Symmetry breaking is seen through minimisation of the
Lagrangian~(\ref{lag}), one solution of which is the vacuum solution
\beq
\Phi(x)=\Phi_0=v\left(\begin{array}{c} 0\\
1\end{array} \right),\ \ \ \ W^\mu=Y^\mu=0.
\eeq
The other minima of~(\ref{lag}) collectively give rise to the vacuum
manifold of degenerate equivalent solutions  
\beq
\label{3sphere}
M = SU(2)_I \x U(1)_Y \cdot \Phi_0 = \{\Phi : \Phi^\dagger \Phi =
v^2\}, 
\eeq
a three-sphere. 

Around the vacuum $\Phi(x)=\Phi_0$ the gauge field mass eigenstates
are given by $W^\mu_1$, $W^\mu_2$, and
\bse
\bea 
\label{gena}
\left( \begin{array}{c} Z^\mu \\ Q^\mu \end{array} \right)
&=& \left( \begin{array}{cc} \cos \theta_w & -\sin \theta_w\\  
             \sin \theta_w & \cos \theta_w \end{array} \right)
\left( \begin{array}{c} W^\mu_3 \\ Y^\mu \end{array} \right),\\
\label{genb}
\left( \begin{array}{c} \al X_Z \\ eX_Q \end{array} \right)
&=& \left( \begin{array}{cc} \cos \theta_w & -\sin \theta_w\\  
             \sin \theta_w & \cos \theta_w \end{array} \right)
\left( \begin{array}{c} \frac{1}{2}ig\si_3 \\ \frac{1}{2}ig'{\bf 1}_2
\end{array} \right), 
\eea
\ese
an orthogonal transformation of the fields, so that the covariant
derivative becomes 
\beq\calD^\mu = \partial^\mu + \sum_{i=1}^{2}\frac{1}{2} ig \si_i W^\mu_i
+ \al X_Z Z^\mu + e X_Q Q^\mu.
\eeq 
Interpreting $Q^\mu$ as the photon constrains $X_Q \Phi_0 =0$,
yielding $\tan \theta_w = g'/g$. This massless gauge field $Q^\mu$ is
associated with the a residual $U(1)_Q$ electromagnetic gauge
symmetry

Explicitly the generators are
\beq
\label{generator}
X_Z = \frac{1}{2}i\left( \begin{array}{cc} \cos 2\theta_w & 0 \\  
             0 & -1 \end{array} \right),\ \ \ \ \ \ 
X_Q = i\left( \begin{array}{cc} 1 & 0 \\ 0 & 0 \end{array} \right), 
\eeq
with $\al = \sqrt{g^2+g'^2}$, $e=gg'/\al$.

The massive gauge field generators $\{\frac{1}{2}i\si_1,
\frac{1}{2}i\si_2, X_Z\}$ form an orthonormal basis for a vector
space $\calM$, such that 
\beq
\label{reddec}
su(2)_I \oplus u(1)_Y = u(1)_Q \oplus \calM,
\eeq
with the orthogonality determined by the inner product
$\inprod{\cdot}{\cdot}$. This relation draws particular comparison to
the isomorphism $M \cong SU(2)_I \x U(1)_Y/U(1)_Q$, associating the
space of massive generators with the tangent space to the vacuum
manifold at $\Phi_0$. 

\section{Vacuum Geometry}

For both the scalar and gauge sectors we now explicitly calculate
their associated metrics. We also calculate the associated
isometry and isotropy groups for each, and from these groups specify the
corresponding geodesic structures.

Having obtained two inequivalent homogenous metrics on the vacuum
manifold we then compare their structure. We relate their isometry and
isotropy groups, and determine when curves are mutually geodesic with
respect to both metrics.

\subsection{Scalar Sector}

The structure associated with the scalar sector is a vector space of
scalar field values ${\bf C}^2$ equipped with a real Euclidean inner
product Re$[\Psi^\dagger \Phi]$. Regarding the vacuum manifold as
embedded within the vector space of scalar field values,
a natural metric may be induced on the vacuum manifold by specifying
its form on each tangent space to be that of the Euclidean inner
product. Such a metric is isotropic and homogenous. Its geodesics are
the great circles. 

Explicitly, regard the tangent space to $M$ at $\Phi \in M$ as an
${\bf R}^3$ subspace of ${\bf C}^2$ 
\beq
\label{tansp}
T_\Phi M = \{\Psi \in {\bf C}^2 : {\rm Re}[\Psi^\dagger \Phi]=0\},
\eeq
with a corresponding metric induced from the real Euclidean inner
product  
\beq
\label{metric-scalar}
g(T_1,T_2)_\Phi = {\rm Re}[T_1^\dagger T_2], \ \ \ T_1, T_2 \in
T_\Phi M. 
\eeq

This metric has an $SU(2)_I \x SU(2)_K$ group of isometries, 
\beq
g(aT_1,aT_2)_{a\Phi}=g(T_1, T_2)_{\Phi},\ \ \ \ 
a \in SU(2)_I \x SU(2)_K. 
\eeq
The $SU(2)_I$ represents the usual
left isospin $SU(2)$ actions upon ${\bf C}^2$ with generators
$\{\frac{1}{2}i \si_a\}$, whilst $SU(2)_K$ acts upon ${\bf C}^2$ with
the generators $\{-\frac{1}{2} \si_2 K, \frac{1}{2} i 
\si_2 K, -\frac{1}{2}i {\bf 1}_2\}$, where $K$ is the complex
conjugation operator. This $SU(2)_K$ contains the hypercharge
$U(1)_Y$, and some other additional symmetries of the scalar sector
which are not symmetries of the full gauge theory.

The isotropy group of $SU(2)_I \x SU(2)_K$ upon $M$ at the point
$\Phi_0$ is the subgroup $SU(2)_{I-K}$ such that 
\beq
SU(2)_{I-K} \cdot \Phi_0 = \Phi_0,
\eeq
which is generated by $\{\frac{1}{2}i \si_1+\frac{1}{2} \si_2 K,
\frac{1}{2}i \si_2-\frac{1}{2} i\si_2 K,\frac{1}{2}i
\si_3+\frac{1}{2}i {\bf 1}_2\}$. This gives the isomorphism
\beq
\label{iso-scalar}
M \cong \frac{SU(2)_I \x SU(2)_K}{SU(2)_{I-K}}.
\eeq
One should be aware that $SU(2) \x SU(2)$ is the compact covering
group of $SO(4)$, just as $SU(2)$ is the compact covering group of
$SO(3)$. Thus Eq.~(\ref{iso-scalar}) is an expression of the more
familiar relation $S^3 \cong SO(4)/SO(3)$.

Given the above isotropy and isometry properties of the metric we can
use the isomorphism~(\ref{iso-scalar}) to calculate the 
geodesics upon the vacuum manifold with respect to the scalar sector
metric $g(\cdot,\cdot)$. This
geodesic structure follows from some results of differential
geometry~\cite[chapter X]{Nomi2}. Specifically, these results
examine the geodesic structure on the coset space, but this may
be simply carried back to $M \subset {\bf C}^2$ to give the results
that we require. We summarise the full result in an appendix and give
only the answer here.

We firstly need an inner product upon $su(2)_I \oplus su(2)_K$ which we
shall define as
\beq
\label{inprodsu2}
\inproda{X_I+X_K}{Y_I+Y_K} = - 2\tr X_IY_I - 2\tr X_KY_K,
\eeq
with an obvious notation. This then induces
\beq
\label{redsu2}
su(2)_I \oplus su(2)_K = su(2)_{I-K} \oplus \calN,
\eeq
where $\calN$ has an orthogonal basis $\{\frac{1}{2}i
\si_1-\frac{1}{2} \si_2 K, \frac{1}{2}i\si_2+\frac{1}{2} i\si_2
K,\frac{1}{2}i \si_3-\frac{1}{2}i {\bf 1}_2\}$. 

The geodesic structure is then 
\begin{quote}\em
the geodesics on $M$ with respect to the metric~$g(\cdot,\cdot)$,
which pass through $\Phi_0$ are: 
\beq
\label{geo-scalar}
\ga_X=\{\exp(Xt)\Phi_0: t \in {\bf R}\},
\eeq
with $X \in \calN$.
\end{quote}

The above geodesic structure is the main result of this
section. Essentially we have derived the geodesic structure to consist
of the great circles embedded in a three-sphere. This result is
as expected, since we are merely
embedding the three-sphere within a Euclidean space. However,
it is the method which is of importance. The same approach may be
adopted for the gauge sector, where the result is intuitively less
clear. Also using the same formalism allows direct comparison between
the metrical structures associated with the gauge and scalar sectors.

We conclude this section by exploring some consequences of the two
Eqs.~(\ref{inprodsu2}, \ref{redsu2}). The space of geodesic generators
$\calN$ is related to the tangent space of $M$ at $\Phi_0$: 
\beq
T_{\Phi_0}M = \calN \cdot \Phi_0.
\eeq
Considering a general $\Phi = a \Phi_0\in M$ with $a \in
SU(2)_I \x SU(2)_K$, there is a more general association
\beq
T_{\Phi}M = \Ad(a)\calN \cdot \Phi.
\eeq
For consistency the Euclidean inner product on $T_{\Phi}M$  
is equivalent to the inner product on $\Ad(a)\calN$ induced by
$\inproda{\cdot}{\cdot}$ of Eq.~(\ref{inprodsu2})
\beq
g(X_1 \Phi, X_2 \Phi)_{\Phi} = \inproda{X_1}{X_2},\ \ \ X_i \in
\Ad(a) \cdot \calN.  
\eeq
This property is essential to the derivation of
Eq.~(\ref{geo-scalar}), and motivates the choice made in
Eq.~(\ref{inprodsu2}).  

\subsection{Gauge Sector}

The main structure associated with the gauge sector is the inner
product $\inprod{\cdot}{\cdot}$ of Eq.~(\ref{inprod}). It specifies
several important related features of the gauge sector. First of all
it defines the gauge kinetic term in Eq.~(\ref{lag}), introducing the
gauge coupling constants as the relative scales. Secondly it
stipulates the embedding of the massive gauge generators $\calM$ in
Eq.~(\ref{reddec}) to be perpendicular to $u(1)_Q$. Finally it renders
the photon $X_Q$, Z-field $X_Z$ and W-field generators mutually
orthonormal.   

We shall use this inner product to define the gauge-sector
metric. The definition is achieved by associating the massive
generators $\calM$ with tangent spaces to the vacuum manifold in a
manner completely analogous to that in the last section. Then the
natural inner product $\inprod{\cdot}{\cdot}$ on the massive
generators $\calM$ induces a metric on the vacuum manifold. We
find the corresponding isometry group of this metric to be
the gauge group $SU(2)_I \x U(1)_Y$ and the isotropy group to be the
residual symmetry $U(1)_Q$. Thus the metric is homogenous, but
anisotropic. Its geodesics are rather complicated in structure.  

Explicitly, observe that the tangent space~(\ref{tansp}) may be 
expressed 
\beq
T_{\Phi_0}M = \calM \cdot \Phi_0. 
\eeq
More generally, the corresponding tangent space at $\Phi=a \Phi_0 \in
M$ is, for any $a \in SU(2)_I \x U(1)_Y$, 
\beq
\label{tgt-iso}
T_\Phi M = aT_{\Phi_0} M = \Ad(a)\calM \cdot \Phi.
\eeq
Transitivity over $M$ guarantees a natural isomorphism between any 
tangent space and some $\Ad(a) \calM$.

Using the isomorphism implied by Eq.~(\ref{tgt-iso}), the inner
product $\inprod{\cdot}{\cdot}$ associates a corresponding
metric on $M$
\beq
\label{metric-gauge}
h(X_1 \Phi, X_2 \Phi)_\Phi = \inprod{X_1}{X_2},\ \ \  X_1, X_2 \in
\Ad(a)\calM. 
\eeq
The precise form is parameterised by the hypercharge
and isospin coupling constants. 

This metric has an $SU(2)_I \x U(1)_Y$ group of isometries
\beq
h(aT_1,aT_2)_{a\Phi}=h(T_1, T_2)_{\Phi},\ \ \ \ 
a \in SU(2)_I \x U(1)_Y. 
\eeq
More precisely, by Eq.~(\ref{tgt-iso}) the action of $a \in SU(2)_I \x
U(1)_Y$ upon $h(\cdot,\cdot)$ is
\beq
h(aT_1,aT_2)_{a\Phi} = \inprod{\Ad(a)X_1}{\Ad(a)X_2} =
\inprod{X_1}{X_2} = h(T_1,T_2)_{\Phi}.
\eeq
The above isometries represent the maximal subgroup of $SU(2)_I \x
SU(2)_K$ leaving $\inprod{\cdot}{\cdot}$ invariant.

%Observe that the isometry group $SU(2)_I \x
%SU(2)_K$ is the maximal compact connected subgroup of $GL({\bf
%C}^2)$. Thus the isometry group of $h(\cdot,\cdot)$ must be contained
%within $SU(2)_I \x SU(2)_K$. Then, by Eq.~(\ref{tgt-iso}) the action
%of $b \in SU(2)_I \x SU(2)_K$ upon $h(\cdot,\cdot)$ is:  
%\bea
%h(bT_1,bT_2)_{b\Phi} = \inprod{\Ad(b)X_1}{\Ad(b)X_2}.
%\eea
%Hence, because $\inprod{\cdot}{\cdot}$ is only invariant upto the adjoint
%action of $SU(2)_I\x U(1)_Y$, the isometry group of $h(\cdot,\cdot)$ is
%$SU(2)_I \x U(1)_Y$.   

The isotropy group of this isometry group at the point
$\Phi_0$ in $M$ is the subgroup $U(1)_Q$ such that 
\beq
U(1)_Q \cdot \Phi_0 = \Phi_0,
\eeq
giving the isomorphism
\beq
\label{iso}
M \cong \frac{SU(2)_I \x U(1)_Y}{U(1)_Q}.
\eeq 
Thus we recover the familiar relation for the vacuum manifold, but
now explicitly associated with the gauge sector metrical structure.

As for the scalar sector the importance of isomorphism~(\ref{iso}) is 
that we may use the isotropy and isometry properties of
the metric to calculate the geodesics upon the vacuum manifold with
respect to the gauge sector metric. Again we give
only answer and refer to the full result summarised in the appendix.  

The structure is 
\begin{quote}\em
the geodesics on $M$ with respect to the metric $h(\cdot,\cdot)$, passing
through $\Phi_0$ are: 
\beq
\label{geo-gauge}
\ga_X=\{\exp(Xt)\Phi_0: t \in {\bf R}\},
\eeq
with $X \in \calM$.
\end{quote}

A short calculation yields the geodesics through $\Phi_0$ associated
with the following generators in $\calM$
\beq
X = \frac{1}{2}i c_1 g\si_1 + \frac{1}{2}ic_2 g\si_2 + d X_Z, \ \ \ \ 
c_1^2 + c_2^2 + d^2 =1
\eeq
to be
\beq
\label{geodesics}
\ga_X(t) = v e^{-id \tan^2 \theta_w t/2} 
\left(\begin{array}{c} c \sin t/2 \\
\cos t/2 - id \sin t/2 \end{array} \right),
\eeq
where $c=c_2 + ic_1$. This structure is rather complicated, for instance
the closed geodesics from a discrete set such that
\bse
\bea
d \tan^2 \theta_w &\in& {\bf Q},\ \ \ \ c \neq 0, \\
d &\in& {\bf Q},\ \ \ \ c =0.
\eea
\ese
Geodesics through other points $\Phi = a \Phi_0 \in M$ may be
simply obtained by acting correspondingly on Eq.~(\ref{geodesics}).

There are a couple of points to observe about the geodesic
structure. Firstly, their exists a totally geodesic two sphere through
$\Phi_0$ defined by those geodesics satisfying $d=0$. Secondly, in the
direction perpendicular to the tangent space of this sphere there
exists a closed geodesic satisfying $c=0$. Because of homogeneity
this is also true for any point on the vacuum manifold. 

The above metric is homogenous but anisotropic. 
Its anisotropy is parametrised by the weak mixing angle
$\theta_w$, becoming isotropic when $\theta_w$ vanishes. Thus one may
interpret the gauge metric $h(\cdot, \cdot)$ as a continuous and
homogenous deformation of the isotropic Euclidean scalar metric
$g(\cdot, \cdot)$. At each point this deformation leaves a two-sphere
and a circle unaffected, but 
deforms relative to the two. In some sense this structure is analogous
to {\em squashing} the three-sphere; having the geometry of an
ellipsoid, but with the deformation homogeneous so as to respect the
homogeneity of~(\ref{iso}). 

We conclude this part of the discussion by observing that the
above structure is rather special to
the three-sphere. It is related to the Hopf fibration. In the Hopf
fibration picture the metric on the $S^1$  
fibres has a different length scale to that on the $S^2$ base
space. One should note that the only other spheres to have a similar
structure are the seven-sphere and the fifteen-sphere.

\subsection{Scalar-Gauge Geometry}

In summary, we found two inequivalent metrics on the electroweak
vacuum manifold 
associated with the scalar and gauge sectors. We shall now determine
how the structure of these metrics relate to each other. 
Comparing the respective symmetry groups determines those symmetries
which are shared. These shared symmetries define two 
submanifolds whose geodesics are mutually geodesic with respect to the
two metrics. These correspond to the two-sphere and the circle
mentioned in the discussion of the gauge sector metric.

The scalar and gauge metrics, $g(\cdot,\cdot)$ and $h(\cdot,\cdot)$,
have the following isometry group decompositions with 
respect to their isotropy groups 
\bse
\bea
su(2)_I \oplus su(2)_K &=& su(2)_{I-K} \oplus \calN,\\
su(2)_I \oplus u(1)_Y &=& u(1)_Q \oplus \calM,
\eea
\ese
where the group structure is related by
\bse
\bea
u(1)_Q &\subset& su(2)_{I-K},\\
u(1)_Y &\subset& su(2)_K,\\
\calM &\subset& su(2)_I \oplus su(2)_K.
\eea
\ese
Also, the tangent space to $M$ at $\Phi_0$ is related to $\calM$ and
$\calN$ by 
\beq
T_{\Phi_0} M = \calM \cdot \Phi_0 = \calN \cdot \Phi_0.
\eeq

It is important to understand how the metrics $g(\cdot,\cdot)$ and
$h(\cdot,\cdot)$ relate to each another. By bilinearity of the
metrics, at the point $\Phi_0 \in M$, 
\beq
g(T_1, T_2)_{\Phi_0} = h(AT_1, AT_2)_{\Phi_0},\ \ \ 
A \in GL(T_{\Phi_0}M),
\eeq
relating the metrics by a linear map of the tangent space. The
eigenspaces of $A$ can be found by explicitly calculating $A$, and
diagonalising it, This yields 
\beq
\label{tdecom}
T_{\Phi_0}M = T_{\Phi_0}^Z M \oplus T_{\Phi_0}^W M,
\eeq
with 
\bse
\bea
T_{\Phi_0}^Z M &=& \calM_Z \cdot \Phi_0,\ \ \ \ \calM_Z = {\bf R}
\cdot X_Z,\\
T_{\Phi_0}^W M &=& \calM_W \cdot \Phi_0,\ \ \ \ \calM_W = {\bf R}
\cdot \frac{1}{2}i\si_1 \oplus {\bf R} \cdot \frac{1}{2}i\si_2. 
\eea
\ese
Then the metrics are related such that
\bea
\label{relation}
g(X_Z\Phi_0 + X_W\Phi_0,Y_Z\Phi_0 + Y_W\Phi_0)_{\Phi_0}&=&\nn \\
\la_Zh(X_Z\Phi_0,Y_Z\Phi_0)_{\Phi_0}  
&+& \la_Wh(X_W\Phi_0,Y_W\Phi_0)_{\Phi_0},
\eea
with an obvious notation, and $\la_Z=\al^2 v^2/\cos 2\theta_w$,
$\la_W=g^2v^2$. This may be easily 
generalised to all $\Phi \in M$ by considering the action of
$SU(2)_I \x U(1)_Y$ on Eq.~(\ref{relation}).

Decomposition~(\ref{tdecom}) also describes the geodesic structure
of $M$ with respect to $g(\cdot,\cdot)$ and $h(\cdot,\cdot)$ in a
rather nice way. Applying Eqs.~(\ref{geo-scalar}) and
(\ref{geo-gauge}), the submanifolds 
\bse
\bea
\label{subM}
M^Z = \exp(\calM_Z)\Phi_0,\ \ \ \ \ T_{\Phi_0}M^Z = T^Z_{\Phi_0} M \\
M^W = \exp(\calM_W)\Phi_0,\ \ \ \ T_{\Phi_0}M^W = T^W_{\Phi_0} M
\eea
\ese
are the {\em only} totally geodesic submanifolds of $M$ with respect
to {\em both} metrics $g(\cdot,\cdot)$ and $h(\cdot,\cdot)$. No other
submanifold of $M$ has this property.

One should note that the $\Ad(U(1)_Q)$-irreducible subspaces
of $\calM$ with respect to the inner product $\inprod{\cdot}{\cdot}$
are 
\beq
\calM = \calM_Z \oplus \calM_W,
\eeq
as found in the decomposition above. These also relate to the mass
eigenstates of the massive gauge fields. This property is in fact very
general~\cite{metom}. 

\section{Physical Implications}

In summary of the previous section we found two homogenous metrics on
the vacuum manifold associated with the scalar and gauge sectors of 
Weinberg-Salam theory. The scalar sector induces an isotropic metric,
whilst the gauge sector induces an anisotropic metric. There is a
unique totally geodesic two sphere with respect to both
metrics. Geodesic curves with respect to both metrics consist of
the geodesics in this two sphere, and one other circular path whose
tangent vector is orthogonal to the tangent plane of this two sphere.

Given this structure one might inquire as to how this relates to the
spectrum of non-perturbative solitonic-type solutions present in
Weinberg-Salam theory. It transpires that the electroweak strings
correspond to the mutually geodesic paths, whilst the
sphaleron corresponds to the mutually geodesic two-sphere. This
approach has the added bonus of interpreting the scattering of
electroweak strings in terms of the holonomy of their respective
geodesic. Also the dynamical stability of the Z-string, as the
weak mixing angle approaches $\pi/2$, is seen to correspond to extreme
anisotropy of the gauge metric.

\subsection{Electroweak Strings}
\label{sec-Electroweak}

Electroweak strings correspond to Nielsen-Olesen vortices embedded in
Weinberg-Salam theory~\cite{Vach92}. As such their boundary conditions
define circular paths on the vacuum manifold. Thus one might expect
that their spectrum and properties should correspond to the geometry
of the vacuum manifold. This is what we find. Their boundary conditions
correspond to the paths that are mutually geodesic with respect to
{\em both} metrics.

Formally, an electroweak string is defined by the embedding
\bea
SU(2)_I \x U(1)_Y &\rightarrow& U(1)_Q\nn \\
\cup \ \ \ &\ & \cup\\
U(1) &\rightarrow& 1,\nn
\eea
with the general Ansatz 
\bse
\bea
\Phi(r, \theta) &=& f_{\rm NO}(r) e^{X\theta}\Phi_0, \\
\ul{A}(r,\theta) &=& \frac{g_{\rm NO}(r)}{r} X \ul{\hat{\theta}},
\eea
\ese
where $X\in su(2)_I \oplus u(1)_Y$ is the vortex generator. 
One may consider only $X \in \calM$, since these minimise the
magnetic energy~\cite{me2}. Thus one considers only
Ans\"atze with boundary conditions geodesic with respect to
the scalar metric $h(\cdot, \cdot)$.

The above vortex Ansatz is a solution provided that~\cite{Barr92}\\
(i) The scalar field must be single-valued. Hence the boundary
conditions describe a closed geodesic with $e^{2\pi
X}\Phi_0=\Phi_0$.\\ 
(ii) The Ansatz is a solution to the equations of motion; then fields
in the vortex do not induce currents perpendicular (in Lie algebra
space) to it~\cite{Vach94}. This may be equivalently phrased
as~\cite{me2}:   
$X$ is a vortex generator if Re$[(X\Phi_0)^\dagger X^\perp \Phi_0] =
0$ for all $X^\perp$ such that $\inprod{X^\perp}{X}=0$.

Condition (ii) can be conveniently restated in terms of the
corresponding metrics:  
\begin{quote}\em 
$X$ is a vortex generator if the associated tangent vector
$T=X\Phi_0$ satisfies $g(T, T^\perp)=0$ for all $T^\perp$ such that
$h(T, T^\perp)=0$. 
\end{quote}
Referring to the discussion around Eq.~(\ref{tdecom}), we see that $T$
must lie in one of the eigenspaces of the linear map relating the two
metrics. Namely $X$ is an element of either $\calM_Z$ or $\calM_W$,
the two $\Ad(U(1)_Q$-irreducible subspaces of $\calM$ in the
decomposition  
\beq
\calM = \calM_Z \oplus \calM_W.
\eeq

It is interesting that the geodesics defined from $\calM_Z$
and $\calM_W$ are the only geodesics which are {\em simultaneously
geodesic with respect to both metrics}. From a mathematical point of
view this is because these geodesics define submanifolds of the vacuum
manifold with coincident scalar and gauge metrics (to an
overall factor). From a physical point of 
view, this may be interpreted as a minimisation of both the scalar and
gauge sectors of the action integral.

From the above it is a fairly trivial exercise to work out the forms
of the vortices, obtaining results in agreement with
refs.~\cite{Vach92, Barr92}.

%: explicitly they are the Z-string
%\bse
%\bea
%\label{zstring}
%\Phi(r,\theta) &=& v f^Z_{\rm NO}(r)
%\left(\begin{array}{c} 0 \\ e^{i \theta} \end{array}
%\right), \\
%\label{zstring2}
%\ul{A}(r,\theta) &=& -\frac{g^Z_{\rm NO}(r)}{r} \cos^2 \theta_w X_Z
%\ul{\hat{\theta}},  
%\eea
%\ese
%and the $W$-strings
%\bse
%\bea
%\label{wstringb}
%\Phi(r,\theta) &=& v f^W_{\rm NO}(r)
%\left(\begin{array}{c} e^{i \eta} \sin \theta \\ \cos \theta
%\end{array} \right),\\
%\label{wstring2}
%\ul{A}(r,\theta) &=& \frac{g^W_{\rm NO}(r)}{r} i(\si_1 \sin \eta +
%\si_2 \cos \eta) \ul{\hat{\theta}},
%\eea
%\ese
%with $\eta$ parameterising the set.

\subsection{Semi-local Vortices}

When the weak mixing angle becomes $\pi/2$ the isospin gauge symmetry
$SU(2)_I$ becomes a global symmetry whilst the hypercharge symmetry
$U(1)_Y$ remains local. Such a model is interpreted as a complex
doublet scalar field with a gauged phase. For suitable scalar
potentials it admits dynamically stable semi-local vortex
solutions~\cite{vach91},
interpreted as the corresponding limit of a Z-string. By continuity
the Z-string is dynamically stable in a region close to weak mixing angle
$\pi/2$~\cite{Vach92}. We wish to point out that this is
related to the gauge metric becoming {\em extremely} anisotropic.

As the weak mixing angle tends to $\pi/2$ the isospin coupling $g$ tends to
zero. For $g=0$, the inner product $\inprod{\cdot}{\cdot}$ becomes ill
defined. It is well defined only on the subalgebra $u(1)_Y$, where it 
takes the form 
\beq
\label{ip1}
\inprod{X}{Y}_{\pi/2} = - \frac{1}{g'^2}\tr X \tr Y.
\eeq
By Eqs.~(\ref{metric-gauge}, \ref{subM}), the gauge metric is only
defined upon the one-dimensional submanifold $M_Z \subset M$, where 
\beq
h_{\pi/2}(X_1 \Phi, X_2 \Phi)_\Phi = \inprod{X_1}{X_2}_{\pi/2},\ \ \ 
X_1, X_2 \in \calM_Z. 
\eeq 
This is interpreted as the limit of {\em extreme} anisotropy; this
anisotropy picking out the submanifold $M_Z \subset M$ upon which the
metric is well defined. 

Physically, the submanifold $M_Z$ represents those points that may be
reached by a gauge transformation from $\Phi_0$. Other points within
the vacuum manifold may only be reached by a global transformation.
This property is related to the stability of the vortex. To decay, the
vortex solution must deform out of $M_Z$. Such a process costs large
gradient energies that may not be compensated for by a gauge
field~\cite{vach91}.  

\subsection{Combination Electroweak Vortices}

When the weak mixing angle vanishes the hypercharge gauge symmetry
$U(1)_Y$ becomes a global symmetry whilst the isospin symmetry
$SU(2)_I$ remains local. Such a model is interpreted as a gauged
complex doublet scalar field with a global phase. Then Weinberg-Salam
theory represents the symmetry breaking $SU(2)_I \rightarrow {\bf 1}$. 
Correspondingly there is a two-parameter family of embedded vortices
representing Z-strings, W-strings, and combination W-Z strings. 
We wish to point out here that this is related to the gauge metric
becoming isotropic, coinciding with the scalar metric and thus all
geodesics of the scalar sector define electroweak strings.

When the weak mixing angle vanishes, so that $g'=0$, the inner
product $\inprod{\cdot}{\cdot}$ becomes well defined only on the
subalgebra $su(2)_I$, 
\beq
\label{ip0}
\inprod{X}{Y}_0 = - \frac{1}{g^2} (2 \tr XY - \tr X \tr Y).
\eeq
Analogous to Eq.~(\ref{metric-gauge}), the metric is then defined from
$\inprod{\cdot}{\cdot}_0$ to be
\beq
h_0(X_1\Phi,X_2\Phi)_\Phi = \inprod{X_1}{X_2}_0,\ \ \  X_1, X_2 \in
su(2)_I. 
\eeq 
Then the scalar and gauge metrics coincide upto a factor of the
isospin coupling constant $g$
\beq
g(T_1,T_2) = \frac{1}{g}h_0(T_1,T_2).
\eeq
Thus the isometry
group of the gauge metric increases to $SU(2)_I \x SU(2)_K$, with this
$SU(2)_K$ now representing a set of global $SU(2)$ symmetries of the
gauge theory only apparent at vanishing weak mixing angle. 

In terms of the electroweak vortex spectrum, condition (ii) is
satisfied for all vortex generators $X \in su(2)_I$, since with
respect to the two metrics all vortices are trivially coincident. 
Hence the spectrum of vortex solutions becomes a continuous family
defined by elements $X \in su(2)_I$.
%\bse
%\bea
%\label{unsqa}
%\Phi(r, \theta)&=& v f_{\rm NO}(r)
%\left(\begin{array}{c} 
%a\sin \theta \\ \cos \theta - ib\sin \theta \end{array} \right),\\
%\label{unsqb}
%\ul{A}(r,\theta) &=& \frac{g_{\rm NO}(r)}{r} (ia_1 \si_1 + ia_2 \si_2
%+ 2b X_Z)\ul{\hat{\theta}}.
%\eea
%\ese
%with $a=a_2+ia_1 \in {\bf C}$ and $b \in {\bf R}$. This family
%consists of $W$-strings, $Z$-strings and combination $WZ$-strings, all
%equivalent under the global $SU(2)_K$ symmetry. 

It is an interesting question to enquire what happens to these
solutions as the weak mixing angle moves off zero. By continuity one
might expect some sort of perturbed solution to exist. However it is
difficult to see what its boundary conditions would be since
necessarily they can only be a geodesic of either the gauge metric or
the scalar metric, not both.

\subsection{Non-Abelian Aharanov-Bohm Scattering}

It is known that the Aharov-Bohm scattering of particles off a vortex,
or one vortex off another, is controlled by the holonomy of a vortex's
boundary conditions~\cite{Wilz}. In this context the term holonomy was
used to indicate non-trivial parallel transport of either a vortex or
a charged particle in a circuit around a vortex. We show here that
this holonomy refers precisely to the holonomy with respect to the
gauge metric.

Associated with the magnetic flux of a vortex is the Wilson line
integral   
\beq
\label{wilson}
U(\theta) = P \exp \left( \int_{0}^{\theta} {\bf A} \cdot d{\bf l}
\right) \subset G, 
\eeq
at infinite radius. The function $U(2\pi)$ dictates the parallel
transport of matter fields around a vortex, such that a fermion
doublet $\Psi$ is transported to $U(2\pi)\Psi$. Diagonalisation of $U$
then associates components $\Psi_i$ with phase shifts $\xi_i$. 
Non-trivial fermionic components $\Psi_i$ interact
with the vortex by an Aharanov-Bohm cross section 
\beq
\deriv{\si}{\theta} = \frac{1}{2\pi k}
\frac{\sin^2(\xi_i/2)}{\sin^2(\theta/2)},
\eeq
whilst trivial components $e^{2 \pi \xi_k}=1$ interact by an Everett
cross section~\cite{Ever}. The above holds for parity symmetric
theories only. When charges for the left and right fermion fields
differ one must also include the effects of induced fermionic zero
modes 

Substitution of the boundary conditions for the Z and W-string,
determined in sec.~(\ref{sec-Electroweak}), gives the Wilson line
integral for electroweak strings. For the Z-string 
\beq
U_Z(\theta) = \exp(-2 X_Z \theta),
\eeq
whilst for the W-string
\beq
U_W(\theta) = \exp(i(\si_1 \sin \eta +
\si_2 \cos \eta) \theta).
\eeq
Vortex boundary conditions restrict $U(2 \pi)\in U(1)_Q$, where
explicitly $U_Z(2\pi)=\exp(-4 \pi \cos^2 \theta_w X_Q)$ and $U_W(2
\pi)={\bf 1}$. For the scattering of a fermion doublet $\Psi$ off a
Z-string this yields phase shifts
$\xi_1=-2 \pi \cos^2 \theta_w$ and $\xi_2=2\pi$. Thus upper components
interact with Z-strings by an Aharanov-Bohm cross section, whilst
lower components interact by an Everett cross section. Fermions
interact with W-strings only by an Everett cross section.  

In \cite{Wilz} the Wilson line integral is related to holonomy,
referring to non-trivial parallel transport around the
vortex. Referring to our appendix, in particular the result
\begin{quote}\em
the parallel transport of ${\bf u} \in T_{\bf v}M$ along
$\tilde{\ga}_X(t)$ is 
${\bf u}' = D(\exp(Xt)){\bf u}.$
\end{quote}
we see that this holonomy refers precisely to the {\em parallel
transport with respect to the gauge sector metric around its closed
geodesics}. For vortices relevant paths are geodesic with
respect to both metrics.

One should be aware that parity violation in the standard model means
that one must take into account the fermionic zero modes. This is done
in ref.~\cite{Gano93}. 

\subsection{The Sphaleron}

Finally, we point out that the existence of the sphaleron solution in
Weinberg-Salam theory is related to the presence an embedded two
sphere in the vacuum manifold that is totally geodesic with respect to
both the gauge and scalar metrics. The dipole moment of a sphaleron is
also related to this embedding.

For $\theta_w=0$ the Ansatz
\bse
\bea
\label{sph-a}
\Phi({\bf r})&=&f_{\rm sph}(r) \exp({\frac{i\pi}{2} \hat{r}_a
\si_a})\Phi_0,\\
\label{sph-b} 
A_a({\bf r}) &=& g_{\rm sph}(r) \frac{i}{2} \epsilon_{abc}\hat{r}_b
\si_c, 
\eea
\ese
constitutes a solution to Weinberg-Salam theory for suitable profile
functions $f_{\rm sph}$, $g_{\rm sph}$. It is unstable because
the boundary conditions define a topologically trivial map $S^2
\rightarrow SU(2)$. By continuity this solution is expected to persist
to non-zero $\theta_w$, with perturbations being produced in the
fields. Such a solution is referred to as the Sphaleron~\cite{Mant}.   

The scalar field asymptotically maps onto the submanifold $M^W \subset
M$. Thus, as with electroweak strings, the boundary conditions
define a totally geodesic submanifold of the vacuum manifold with
respect to {\em both} the scalar and gauge metrics. One should note
that $M^W$ is the only two-dimensional submanifold of the vacuum
manifold defined so. 

To prevent an electric monopole component at non-zero weak mixing angle,
the gauge field (\ref{sph-b}) deforms such that $\frac{i}{2}\si_3
\rightarrow X_Z$. This implies a preferred axis in the
$\rhat^3$-direction, with the configuration rotationally symmetric
about it. Inducing, to lowest order, dipolar perturbations
in the electromagnetic field  
\beq
\label{dipole}
\de Q_a = \frac{\epsilon_{abc} \mu_b \hat{r}_c}{4 \pi r^3} \/ X_Q,
\eeq
with $\ul{\mu}$ parallel to $\rhat^3$, as found by substitution into
the field equations~\cite{Mant}. 

\section{Discussion}

In this final section we briefly discuss some extensions to this work
and make some comments that may warrant further note.
\\
(i) {\bf The General Case}\\
The group theory in this paper can be extended to the general symmetry
breaking $G \rightarrow H$ in a fairly straightforward manner. Hence,
in general, one may expect two metrics on the vacuum manifold relating
to the scalar and gauge sector. As in Weinberg-Salam theory,
embedded vortices will be geodesic with with respect to both metrics,
and the Aharonov-Bohm scattering will relate to the holonomy of the
gauge sector metric.\\
(ii) {\bf Simplicity of Electroweak Theory}\\
Crucial to establishing the scalar and gauge metrics on the vacuum
manifold was establishing the isomorphisms with the
coset spaces $SU(2)_K \x SU(2)_I / SU(2)_{I-K}$ and $SU(2)_I \x
U(1)_Y/U(1)_Q$. The first of which is a symmetric space, and the
second is a non-symmetric homogenous space, interpreted as deformed
from the first. This structure is in fact quite special, and
electroweak theory constitutes the smallest dimensional example of
this. \\

\noindent
(iii) {\bf Energetics and Curvature}\\
From the metrical structure one has a corresponding curvature of the
vacuum manifold. It seems sensible that the energy of embedded
vortices should be associated with the sum of the curvatures of
the scalar and gauge metrics on the submanifold of the vacuum manifold
associated with the scalar boundary conditions of the
vortex. Coefficients of this sum should be related to those of the
Landau potential, and the value of this sum should be related to the
stability of the vortex.\\ 
(iv) {\bf Insensitivity to the Form of the Theory}\\
In relating the solitonic spectrum and properties to the metrical
structure of the theory one moves away from the specific details of
the Lagrangian. Thus the metrical approach relates more to the general
symmetry features of the theory rather than the specific model of
symmetry breaking.

%Within a related paper~\cite{me} the problem of squashing
%general symmetric spaces is addressed. Dealing specifically with
%continuously and homogenously squashing the metric on a symmetric
%space the following result is found:
%\begin{result}
%The only symmetric spaces whose metrics may be continuously and
%homogenously squashed are the Lie groups $I$. Then the squashed
%metrics are defined by the Cartan subgroup $K_I \subset I$ on the
%topologically isomorphic homogenous space 
%\beq
%I_{\rm squash} = \frac{I \x K}{K_{\rm diag}},
%\eeq 
%where $K \cong K_I$, and $K_{\rm diag}$ is the diagonal subgroup of
%$K$ and $K_I$. 
%\end{result}
%A way of interpreting this is that only symmetric spaces that may be
%endowed with a group structure may be squashed. Then the squashing is
%specified by the Cartan subgroup of that manifold, which is squashed
%relative to the overall manifold. For instance, the electroweak vacuum
%is squashed along the $S^1_Z$ fibre, defined by the Cartan subgroup of
%$SU(2)$; namely $U(1)$. 

\bigskip

%%%%%%%%%%%%%%%%%%%%%%%%%%%%% acknowledgements %%%%%%%%%%%%%%%%%%%

{\noindent{\Large{\bf Acknowledgements.}}}
\nopagebreak
\bigskip
\nopagebreak

This work was supported in part by PPARC.  N.L. acknowledges
EPSRC for a research studentship and King's College, Cambridge for 
a junior research fellowship. This work was supported in part by 
the European Commission under the Human Capital and Mobility program,
contract no. CHRX-CT94-0423.

\bigskip
\bigskip

%%%%%%%%%%%%%%%%%%%%%%%%% appendix  %%%%%%%%%%%%%%%%%%%%%%%%%

{\noindent{\Large{\bf Appendix}}} 
\par
\nopagebreak
\bigskip
\nopagebreak
\noindent
We provide here a quick summary of the results of \cite[chapter
X]{Nomi2} that are relevant to this work.

Consider a manifold $G/H$, where $H \subset G$ are compact Lie
groups. Then an important and relevant decomposition is the reductive
decomposition of the Lie algebras
\beq
\calG = \calH \oplus \calM,
\eeq
which satisfies
\bse
\bea
\Ad(H) \calH \subseteq \calH,\\
\Ad(H) \calM \subseteq \calM.
\eea
\ese
Here $\calM$ may be associated with the tangent space to $M$ at the
trivial coset. One should note that strictly speaking an inner product
$\inprod{\cdot}{\cdot}$ on $\calG$ is required to define the reductive
decomposition.  

Given such a decomposition we can associate a $G$-invariant connection
on $G/H$ having the following properties:\\
(i)\ \ All geodesics in $M$ emanating from the trivial coset are of
the form
\beq
\ga_X(t) = \exp(Xt)H,
\eeq
with $X \in \calM$.\\
(ii)\ \ Considering $\exp(Xt)\subset G$, the parallel transport of a
tangent vector $Y \in \calM$ along the curve $\exp(Xt)H$ is
\beq
Y' = \Ad(\exp(Xt))Y.
\eeq 
(iii)\ \ The corresponding $G$-invariant metric on $G/H$ is associated
with the inner products on $\calG$ such that at the trivial coset
\beq
g(X, Y)_H = \inprod{X}{Y},
\eeq
identifying tangent vectors with elements of $\calM$.

By applying the above result one may examine the case with a vector
space $V$ such that $G$ acts on it by the $D$-representation. Then
$H_{\bf v}$ is the isotropy subgroup at ${\bf v} \in V$, and we
associate the manifold 
\beq
M = D(G){\bf v} \cong \frac{G}{H}.
\eeq
Given an inner product $\inprod{\cdot}{\cdot}$ on $\calG$ we 
associate a $G$-invariant connection on $M$ from the decomposition
$\calG = \calH_{\bf v} \oplus \calM$. Denoting the tangent space at
${\bf v}$ to $M$ by $T_{\bf v}M = \calM {\bf v}$, the
following properties are apparent\\
(i)\ \ All geodesics emanating from ${\bf v}$ are of the form
\beq
\tilde{\ga}_X(t) = D(\exp(Xt)){\bf v}.\\
\eeq
(ii) The parallel transport of ${\bf u} \in T_{\bf v}M$ along
$\tilde{\ga}_X(t)$ is 
\beq
{\bf u}' = D(\exp(Xt)){\bf u}.
\eeq
(iii)\ \ The corresponding $G$-invariant metric on $M$ takes the
value at ${\bf v}$
\beq
g(X{\bf v}, Y{\bf v})_{\bf v} = \inprod{X}{Y}.
\eeq

%%%%%%%%%%%%%%%%%%%%%%%%%%%  end of text   %%%%%%%%%%%%%%%%%%%%%%%

%bibliography

%end of document
\end{document}